\documentclass[hyper,letterpaper,12pt]{article}
\usepackage{a4wide}
\usepackage{amsmath}
\usepackage[
      colorlinks=true,
      linkcolor=blue,
      urlcolor=blue,
      filecolor=black,
      citecolor=blue,
            ]{hyperref}
\usepackage{color}
\usepackage{graphicx}
\usepackage{amsfonts}
\usepackage{amssymb}
\usepackage{epstopdf}

\def\beq{\begin{equation}}
\def\eeq{\end{equation}}

\renewcommand{\(}{\left(}

\renewcommand{\[}{\left[}
\renewcommand{\]}{\right]}

\begin{document}
\title{\bf \Large Entanglement entropy in a holographic p-wave superconductor model}

\author{\large
~~Li-Fang Li$^1$\footnote{E-mail: lilf@itp.ac.cn}~,
~Rong-Gen Cai$^2$\footnote{E-mail: cairg@itp.ac.cn}~,
~~Li Li$^2$\footnote{E-mail: liliphy@itp.ac.cn}~,
~~Chao Shen$^1$\footnote{E-mail: sc@nssc.ac.cn}
\\
\\
\small $^1$State Key Laboratory of Space Weather, \\
\small Center for Space Science and Applied Research, Chinese Academy of Sciences,\\
\small Beijing 100190, China.\\
\small $^2$State Key Laboratory of Theoretical Physics,\\
\small Institute of Theoretical Physics, Chinese Academy of Sciences,\\
\small Beijing 100190,  China.
}
\date{\today}
\maketitle

\begin{abstract}
\normalsize In a recent paper, arXiv:1309.4877, a holographic p-wave model has been proposed in an Einstein-Maxwell-complex vector field theory with a negative cosmological constant. The model exhibits rich phase structure depending on the mass and the charge of the vector field. We investigate the behavior of the entanglement entropy of dual field theory in this model. When the above two model parameters change, we observe the second order, first order and zeroth order phase transitions from the behavior of the entanglement entropy at some intermediate temperatures. These imply that the entanglement entropy can indicate not only the occurrence of the phase transition, but also the order of the phase transition. The entanglement entropy is indeed a good probe to phase transition. Furthermore, the ``retrograde condensation"  which is a sub-dominated phase is also reflected on the entanglement entropy.
\end{abstract}

\tableofcontents

\section{ Introduction}
\label{sect:introduction}
Over the past years, the AdS/CFT correspondence~\cite{Maldacena:1997re,Gubser:1998bc,Witten:1998qj} has been
extensively applied to the study of superconductor/superfluid in condensed matter physics. For reviews, please see refs.~\cite{Hartnoll:2009sz,Herzog:2009xv,McGreevy:2009xe,Horowitz:2010gk,Cai:2015cya}. Recently
we have proposed a holographic model of p-wave superconductors
in refs.~\cite{Cai:2013pda,Cai:2013aca} by introducing a complex vector
field $\rho_\mu$ charged under a Maxwell gauge field $A_\mu$ in the AdS
bulk. This model contains two parameters, the mass and charge of the vector field. Depending on the
model parameters, it has been shown that this model exhibits a rich phase structure~\cite{Cai:2013aca,Cai:2014ija}.
In this model, the second order, first order and zeroth order phase
transitions may happen.  We also found the so-called ``retrograde condensation", but it is
sub-dominated in the sense that its free energy is much larger than
the black hole solution without the vector hair. Comparing the holographic p-wave model by introducing a SU(2) gauge field in the bulk proposed in ref.~\cite{Gubser:2008wv}, it turns out that
the new p-wave model is a generalization of the one in ref.~\cite{Gubser:2008wv} in the sense that the vector field has a general mass
and gyromagnetic ratio~\cite{Cai:2013kaa}.

The entanglement entropy is a good probe to quantum phase transitions and various phases in
quantum field theory and many-body system. The holographic entanglement
entropy proposed in ref.~\cite{Ryu:2006bv} provides an effective way to calculate the
entanglement entropy for a strongly coupled field theory (for reviews see refs.~\cite{Nishioka:2009un,Takayanagi:2012kg}).
The authors of refs.~\cite{Albash:2012pd,Cai:2012nm,Arias:2012py,Peng:2014ira,Yao:2014fwa} studied the behaviors of
entanglement entropy in holographic conductor/superconductor phase transition, including
s-wave and p-wave cases. The results showed that the entanglement entropy decreases
as one lowers temperatures, indicating the degrees of freedom are reduced at lower temperature.
And the behavior of the entanglement entropy changes dramatically when the order
of the phase transition changes, which implies that the entanglement entropy is indeed
a good probe to phase transition. The behaviors of entanglement entropy for
the holographic s-wave and p-wave superconductor/insulator model at zero temperature have also been investigated
in refs.~\cite{Cai:2012sk,Cai:2012es,Yao:2014fua,Cai:2013oma}.  This case is very similar to the study of confinement/deconfinement
phase transition by the holographic entanglement entropy~\cite{Nishioka:2006gr,Klebanov:2007ws}.
Unlike the conductor/superconductor phase transition, the
entanglement entropy in the s-wave superconductor/insulator case~\cite{Cai:2012sk,Cai:2012es,Yao:2014fua} as a function
of chemical potential is not monotonic: at the beginning of the transition, the entropy
first increases and reaches its maximum at some chemical potential, and then decreases
monotonically.

In this paper, we will study the entanglement entropy in this new holographic p-wave superconductor model~\cite{Cai:2013pda,Cai:2013aca}. The entanglement entropy is calculated for a straight strip geometry with the holographic proposal. We find that the behavior of entanglement entropy changes qualitatively in different phases as we alter the mass square $m^2$ and the charge $q$ of the vector field. Our results show that the entanglement entropy can capture the characteristics of phase transitions and various phases in the holographic superconductor model. By comparing the entanglement entropy with the thermal entropy of the bulk black holes during the whole process of phase transition, we see the possibility for understanding the black hole entropy as the entanglement entropy~\cite{Jacobson:1994iw,Kabat:1995eq,Solodukhin:2006xv,Emparan:2006ni}. Furthermore, the ``retrograde condensation" which is a sub-dominated phase is also reflected on the entanglement entropy.

This paper is organized as follows. In the next section, we briefly review the holographic p-wave superconductor model. The whole system with full back raection is solved by shooting method and the main phase structure of the model is briefly summarized. Section~\ref{sect:entanglement_entropy} is devoted to exploring the behaviors of the entanglement entropy in the p-wave superconductor model. We present numerical results in this section and for each given $m^2$, we scan a wide range of $q$ to find all possible behaviors of the entanglement entropy. The conclusions and some discussions are included in section~\ref{sect:conclusion}.


\section{The holographic model}
\label{sect:model}

Let us start with the holographic model of the p-wave superconductors proposed in refs.~\cite{Cai:2013pda,Cai:2013aca}
\begin{equation}\label{action}
S=\frac{1}{2\kappa^2}\int d^4 x
\sqrt{-g}\left(\mathcal{R}+\frac{6}{L^2}-\frac{1}{4}F_{\mu\nu} F^{\mu
\nu}-\frac{1}{2}\rho_{\mu\nu}^\dagger\rho^{\mu\nu}-m^2\rho_\mu^\dagger\rho^\mu\right),
\end{equation}
where $\kappa^2\equiv 8\pi G $ with $G$ the gravitational constant,
$L$ is the radius of AdS spacetime. $F_{\mu\nu}$ and $\rho_{\mu}$ are Maxwell
field and complex vector field, respectively, $F_{\mu\nu}=\nabla_\mu
A_\nu-\nabla_\nu A_\mu$ with $\nabla$ the covariant derivative
associated with the metric $g_{\mu\nu}$,
$\rho_{\mu\nu}=D_\mu\rho_\nu-D_\nu\rho_\mu$ with $D_\mu=\nabla_\mu-iq A_\mu$. And $m^2$ and $q$ are the mass square and
the charge of the complex vector field, respectively. Varying the action, we have the equations of motion for gauge field $A_{\mu}$ and charged vector field $\rho_{\mu}$ as
\begin{eqnarray}\label{gauge}
\nabla^\nu F_{\nu\mu}-iq(\rho^\nu\rho_{\nu\mu}^\dagger-{\rho^\nu}^\dagger\rho_{\nu\mu})&=&0,\nonumber\\
D^\nu\rho_{\nu\mu}-m^2\rho_\mu&=&0,
\end{eqnarray}
and the corresponding Einstein's field equations
\begin{equation}\label{tensor}
\begin{split}
\mathcal{R}_{\mu\nu}-\frac{1}{2}\mathcal{R}g_{\mu\nu}&-\frac{3}{L^2}g_{\mu\nu}=\frac{1}{2}F_{\mu\lambda}{F_\nu}^\lambda+\frac{1}{2}\mathcal{L}_m g_{\mu\nu}\\
&+\frac{1}{2}\[(\rho_{\mu\lambda}^\dagger{\rho_\nu}^\lambda+m^2{\rho_\mu}^\dagger\rho_\nu)+\mu\leftrightarrow\nu\].
\end{split}
\end{equation}
\label{sect:motion}
The system admits an analytical solution with vanishing $\rho_\mu$, corresponding to the normal phase, which is just the planar AdS Reissner-Nordstr\"om black hole given by~\cite{Cai:1996eg}
\begin{equation}\label{RN}
\begin{split}
ds^2=-A(r)dt^2+\frac{dr^2}{A(r)}+r^2(dx^2+dy^2),\\
A(r)=r^2-\frac{1}{r}(r_h^3+\frac{\mu^2 r_h}{4})+\frac{\mu^2 r_h^2}{4r^2}, \quad \phi(r)=\mu(1-\frac{r_h}{r}),
\end{split}
\end{equation}
with $r_h$ is the horizon radius and $\mu$ is the chemical potential of the dual field theory. The Hawking temperature of the black hole is $T=\frac{r_h}{4\pi}(3-\frac{\mu^2}{4r_h^2})$.

When we tune the temperature, the system exhibits an instability which triggers the condensation of the charged vector field $\rho_{\mu}$. Here we consider the following ansatz~\cite{Cai:2013aca}
\begin{equation}\label{ansatz}
\begin{split}
ds^2=-f(r)e^{-\chi(r)}dt^2+\frac{dr^2}{f(r)}+r^2h(r)dx^2+r^2dy^2,\\
\rho_\nu dx^\nu=\rho_x(r)dx,\quad A_\nu dx^\nu=\phi(r)dt.
\end{split}
\end{equation}
The temperature $T$ of the black hole is given by
\begin{equation}\label{temp}
T=\frac{f'(r_h)e^{-\chi(r_h)/2}}{4\pi}.
\end{equation}
And the thermal entropy of the black hole is given by the Bekenstein-Hawking area formula: $S= 2\pi V_2 r_h^2/\kappa^2$, where $V_2$ is the area spanned by coordinates $x$ and $y$. According to the AdS/CFT correspondence, the solution~\eqref{RN} with vanishing $\rho_\nu$ corresponds
to a conductor (normal) phase, while the solution~\eqref{ansatz} with the complex vector hair
$\rho_\nu$ corresponds to the superconducting phase of the dual field theory. The process from the black hole without hair to the black hole with non-trivial vector hair mimics the conductor/superconductor phase transition. To study the behavior of the entanglement entropy during this process, we need to firstly obtain the hairy black hole solution.

With the ansatz~\eqref{ansatz}, the independent equations of motion turn out to be
\begin{equation}\label{eoms}
\begin{split}
\phi''+(\frac{h'}{2h}+\frac{\chi'}{2}+\frac{2}{r})\phi'-\frac{2q^2\rho_x^2}{r^2fh}\phi=0,\\
\rho_x''+(\frac{f'}{f}-\frac{h'}{2h}-\frac{\chi'}{2})\rho_x'+\frac{e^{\chi}q^2\phi^2}{f^2}\rho_x-\frac{m^2}{f}\rho_x=0, \\
\chi'-\frac{2f'}{f}-\frac{h'}{h}+\frac{\rho_x'^2}{rh}-\frac{re^\chi\phi'^2}{2f}-\frac{e^\chi q^2\rho_x^2\phi^2}{rf^2h}+\frac{6r}{L^2f}-\frac{2}{r}=0,\\
h''+(\frac{f'}{f}-\frac{h'}{2h}-\frac{\chi'}{2}+\frac{2}{r})h'+\frac{2{\rho_x'}^2}{r^2}-\frac{2e^\chi q^2\rho_x^2\phi^2}{r^2f^2}+\frac{2m^2\rho_x^2}{r^2f}=0,\\
(\frac{2}{r}-\frac{h'}{2h})\frac{f'}{f}+(\frac{1}{r}+\frac{\chi'}{2})\frac{h'}{h}-\frac{\rho_x'^2}{r^2h}+\frac{e^\chi\phi'^2}{2f}+\frac{3e^\chi q^2\rho_x^2\phi^2}{r^2f^2h}-\frac{m^2\rho_x^2}{r^2fh}-\frac{6}{L^2f}+\frac{2}{r^2}=0,
\end{split}
\end{equation}
where the prime denotes the derivative with respect to $r$. We will use shooting method to solve the above equations~\eqref{eoms}. To do this, we have to first specify the boundary conditions
for this system. Near the AdS boundary $ r \to \infty$, these fields have the following expansion behaviors:
\begin{equation} \label{boundary}
\begin{split}
\phi=\mu-\frac{\rho}{r}+\ldots,\quad \rho_x=\frac{{\rho_x}_-}{r^{{\Delta}_-}}+\frac{{\rho_x}_{+}}{r^{{\Delta}_+}}+\ldots,\\
f=r^2(1+\frac{f_3}{r^3})+\ldots,\quad h=1+\frac{h_3}{r^3}+\ldots,\quad \chi=0+\frac{\chi_3}{r^3}+\ldots,
\end{split}
\end{equation}
with ${\Delta}_\pm=\frac{1\pm\sqrt{1+4 m^2}}{2}$. According to the AdS/CFT dictionary, the constants $\mu$ and $\rho$ can be interpreted as the chemical potential and the charge density in the dual field theory, respectively. ${\rho_x}_-$ is the source of the dual operator and ${\rho_x}_+$ gives its expectation value. To require the U(1) symmetry being broken spontaneously, we will take ${\rho_x}_-=0$ in the numerical calculation.

We impose regular conditions on the horizon $r=r_h$. Concretely, one has $f(r_h)=0$ and $\phi(r_h)=0$. Then we are left with five independent parameters $\{r_h,\rho_x(r_h),\phi'(r_h),h(r_h),\chi(r_h)\}$. The scaling symmetry
\begin{equation} \label{scaling1}
e^\chi\rightarrow\lambda^2 e^\chi,\quad t\rightarrow\lambda t,\quad \phi\rightarrow\lambda^{-1}\phi,
\end{equation}
can be used to set $r_h=1$ for performing numerics. Similarly, the scaling symmetries

\begin{equation} \label{scaling2}
\rho_x\rightarrow\lambda \rho_x,\quad x\rightarrow\lambda^{-1} x,\quad h\rightarrow\lambda^2 h,
\end{equation}
and
\begin{equation} \label{scaling3}
r\rightarrow\lambda r,\quad (t,x,y)\rightarrow{\lambda^{-1}}(t,x,y),\quad(\phi,\rho_x)\rightarrow\lambda(\phi,\rho_x), \quad f\rightarrow\lambda^2f,
\end{equation}
lead us to set $\{\chi(r_h)=0, h(r_h)=1\}$. Thus we finally have two independent parameters $\{\rho_x(r_h),\phi'(r_h)\}$ at hand. Given $\phi'(r_h)$ as the shooting parameter to match the source free condition, i.e., ${\rho_x}_-=0$, we can solve the set of the fully coupled differential equations. The details of numerical calculation can be found in ref.~\cite{Cai:2013aca}.

In order to determine which phase is thermodynamically favored, we should calculate the free energy of the system for both normal phase and condensed phase. We will work in grand canonical ensemble with fixed chemical potential. Following ref.~\cite{Cai:2013aca}, the free energy $\Omega$ can be expressed as
\begin{equation}\label{free}
\Omega=\frac{V_2}{2\kappa^2}(f_3+h_3-\chi_3).
\end{equation}
For the normal phase given in~\eqref{RN}, one has $f_3=-r_h^3-\frac{\mu^2 r_h}{4}$, and $h_3=\chi_3=0$.

We have studied the phase structure of the model by investigating thermodynamics of various solutions of the system. There exists a critical
mass square $m^2_c = 0$. The system shows qualitatively different properties for the cases $m^2 \ge m^2_c$  and $m^2< m^2_c$.

\begin{figure}[h]
\centering
\includegraphics[scale=0.92]{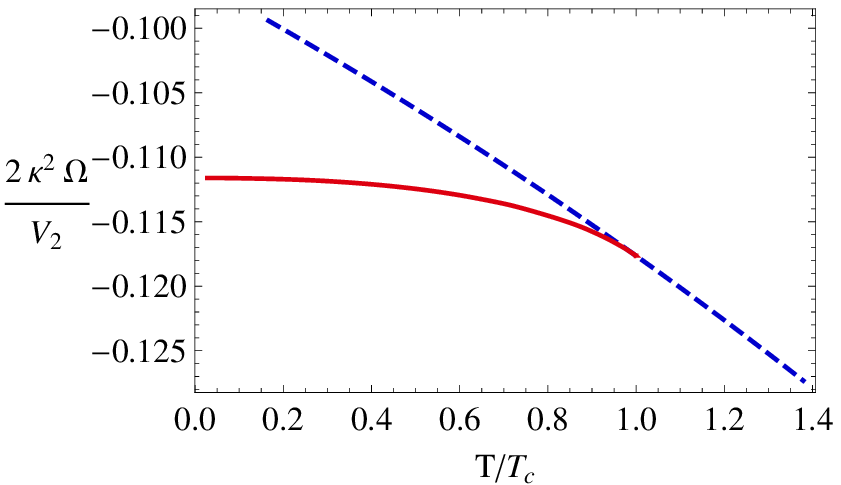}\ \ \ \
\includegraphics[scale=0.92]{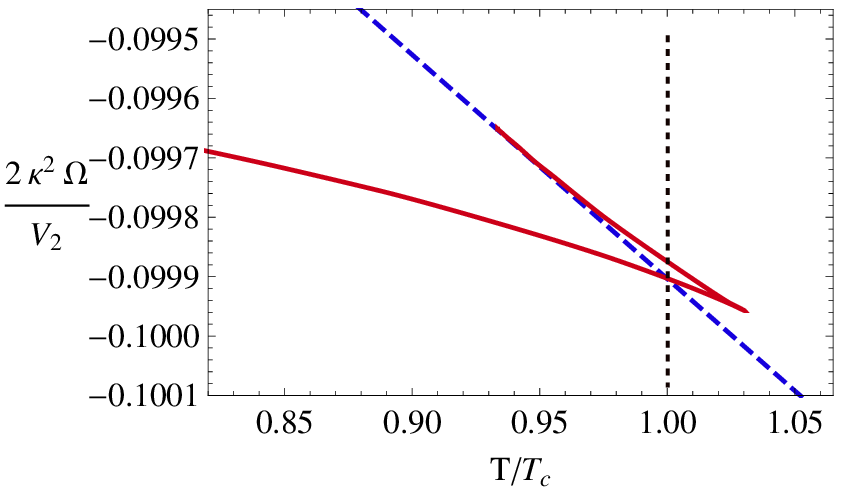} \caption{\label{freep} The grand potential $\Omega$ as a function of temperature in the case $m^2=3/4$ for $q=3/2$~(left plot) and $q=6/5$~(right plot). The critical temperature $T_c$ marks the occurrence of the superconducting phase. Trace the physical curve by choosing the lowest grand potential at a fixed T. The $q=3/2$ plot shows the typical second-order phase transition. The $q=6/5$ plot disposes a first order phase transition.}
\end{figure}

When $m^2 \ge m_c^2$,  there exists a critical charge $q_c$, which depends on the value of $m^2$. If $q\ge q_c$, the system undergoes
a second order phase transition at the critical temperature $T=T_c$ from the conductor (normal) phase for temperature $ T >T_c$ to the
superconducting phase for temperature $ T<T_c$. On the other hand, if $q<q_c$, the phase transition becomes a first order one, which can be seen from figure~\ref{freep}.

\begin{figure}[h]
\centering
\includegraphics[scale=0.92]{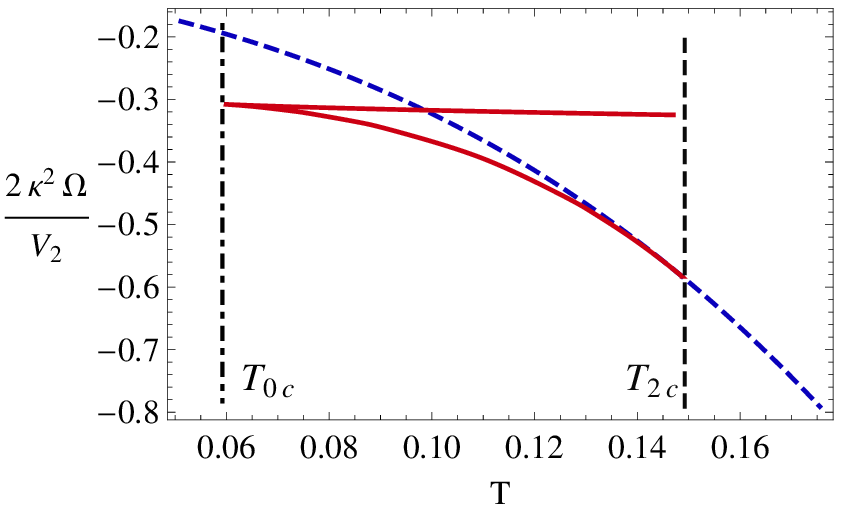}\ \ \ \
\includegraphics[scale=0.92]{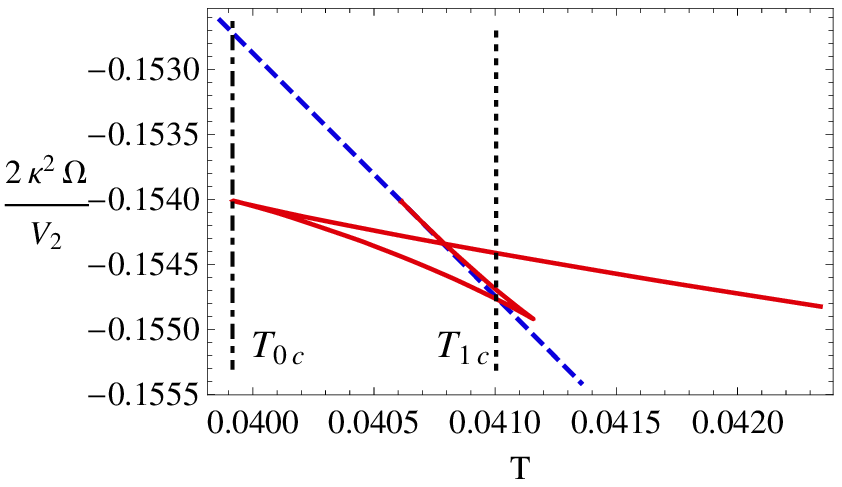}\\
\includegraphics[scale=0.92]{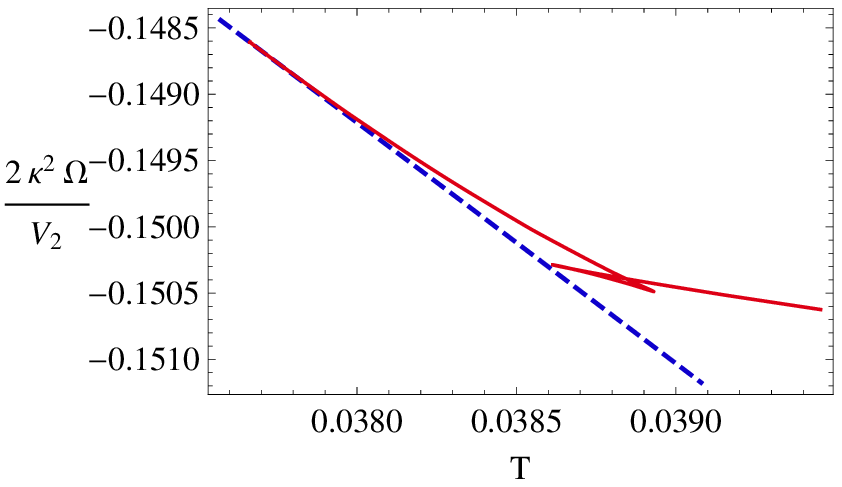}\ \ \ \
\includegraphics[scale=0.92]{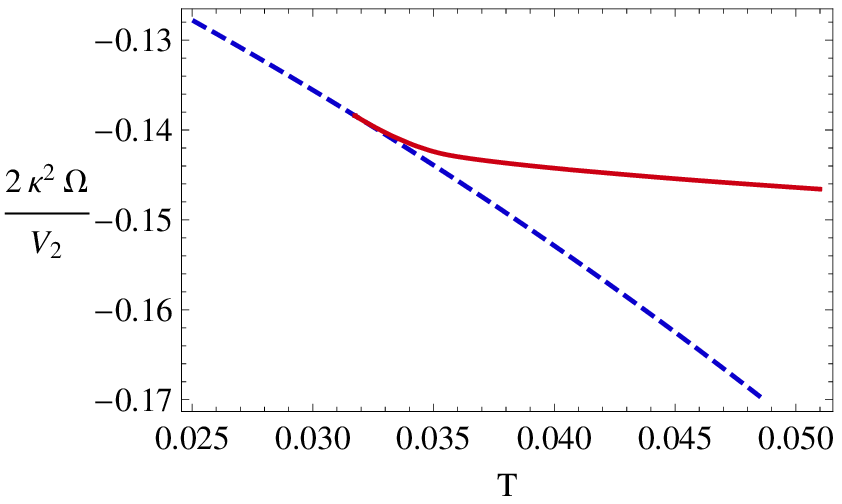}
 \caption{\label{freen} The free energy $\Omega$ as a function of temperature with vector condensation (solid red), and without the vector condensation (dashed blue) in the cases $m^2=-3/16$ and $q=2$ (top left), $q=39/40$ (top right), $q=19/20$ (bottom left), $q=9/10$ (bottom right). The $q=2$ plot shows the second order phase transition but the condensed phase terminates at finite low temperature $T_0$. The $q=39/40$ plot exhibits a first order phase transition but also the superconducing phase terminates at finite low temperature $T_0$. The $q=19/20$ and $q=9/10$ plots demonstrate that the condensed phase has free energy much larger than  the normal phase and thus is not thermodynamical favored.}
\end{figure}

When $m^2 < m_c^2$, the range of charge $q$ is divided into three regimes by two critical charges, $q_{\alpha}$ and $q_{\beta}$, which also depends on
the value of $m^2$. If $q \ge q_{\alpha}$, the system is in a normal phase when $T >T_{2c}$, in a superconducting phase when $ T_{0c} <T <T_{2c}$, and in the normal phase when $ T<T_{0c}$. There exist a second order phase transition at $T=T_{2c}$, where the free energy is continuous,  and a zeroth order phase transition at $T=T_{0c}$, where the free energy is discontinuous. If $q_{\beta} < q < q_{\alpha}$, the
second order phase transition is replaced by a first order one at $T=T_{1c}$.  If $q \le q_{\beta}$, the black hole solution with vector hair can exist, but its free energy is always larger than the one without the vector hair. The condensation is called ``retrograde condensation". The solution is sub-dominated and therefore this phase will not appear in the phase diagram. In this case, the system is always in the normal
phase. We can see these clearly from figure~\ref{freen}. Note that all the critical temperatures $T_c$, $T_{2c}$, $T_{1c}$ and $T_{0c}$ depend on the model parameters $m^2$ and $q$.

In summary, depending on the model parameters $m^2$ and $q$, the model exhibits a rich phase structure. The details are summarized in Table~\ref{tab:1}. In the spirit of AdS/CFT correspondence, the mass square $m^2$ of the vector field is related to the conformal dimension of the dual vector operator, while the charge $q$ of the vector field corresponds to the ``charge" of the condensate. From the bulk point of view the charge $q$ is the coupling parameter between the vector field and the bulk Maxwell field and governs the back reaction strength of the matter fields on the background geometry. In this sense the charge $q$ plays the same role as the coupling parameter $\hat{g}^2$ in the Einstein-Yang-Mills theory~\cite{Ammon:2009xh}. Thus one may think that the parameter $q$ counts the additional degrees of freedom in the dual CFT~\cite{Ammon:2009xh}. In this way the mass square and charge of the vector filed control the phase structure of the dual CFT.


%
\begin{table}
  \centering
  \begin{tabular}{|c|c|c|c|c|c|}
    \hline
    \small{Charge} & \multicolumn{5}{|c|}{Phase transition and its order for $m^2 \ge m_c^2$}\\
    \hline
    $q \ge q_c$ & \multicolumn{2}{|c|}{$T>T_{c}$, N} & $T=T_{c}$, $2^{nd}$  & \multicolumn{2}{|c|}{$T<T_{c}$, SC} \\
    \hline
    $q < q_c$&\multicolumn{2}{|c|}{$T>T_{c}$, N }& $T=T_{c}$, $1^{st}$ & \multicolumn{2}{|c|}{$T<T_{c}$, SC}\\
   \hline
      &\multicolumn{5}{|c|}{ Phase transition and its order for $ m^2 < m^2_{c}$} \\
    \hline
    $q \ge q_{\alpha}$ & $T>T_{2c}$, N & $T=T_{2c},2^{nd}$ & $T_{0c}<T<T_{2c}$, SC & $T=T_{0c}, 0^{th}$ & $T<T_{0c}$, N \\
    \hline
    $q_{\beta}< q<q_{\alpha}$ &$T>T_{1c}$, N & $T=T_{1c},1^{st}$ & $T_{0c}<T<T_{1c}$, SC & $T=T_{0c}, 0^{th}$ & $T<T_{0c}$, N\\
    \hline
    $q \le q_{\beta}$ & \multicolumn{5}{|c|}{N}  \\
    \hline
  \end{tabular}
  \caption{The phase transition and its order with respect to the charge $q$ and the mass square $m^2$. Here N stands for conductor (normal) phase with RN-AdS black hole solution, while SC for the superconducting phase with black hole solutions with vector hair. $0^{th}$, $1^{st}$, and $2^{nd}$ stand for the zeroth order, first order and second order phase transition, respectively.
  $m_c^2=0$. $q_c$, $q_\alpha$, $q_\beta$ are some critical values depending on $m^2$. $T_c$, $T_{0c}$, $T_{1c}$, and $T_{2c}$ are some critical temperatures depending on $m^2$ and $q$.}
  \label{tab:1}
\end{table}

In addition, let us mention here that the zeroth order phase transition with discontinuous free energy looks
strange at first glance, but it has been argued that it could appear in superconductivity and superfluid~\cite{vp}.
In that reference an exact solvable model has  also been proposed. Furthermore, the zeroth order phase transition also appears
in the extended phase space of AdS black hole thermodynamics~\cite{Altamirano:2013ane,Gunasekaran:2012dq}.  But we remind
here that the zeroth order phase transition might not be physical in this holographic superconductor model.
 The reason is that in this model we have only considered
two phases, the normal phase described by the RN-AdS black hole and the superconducting phase described by black hole with
vector hair, both solutions are of the translation symmetry in the transversal directions. If  further consider other components
of the complex vector field and/or the case without the transversal translation symmetry, we might find other phases which have
a lower free energy so that those phases are more stable and the zeroth order phase transition discussed in the above will
be replaced by a first order or second order phase transition.  But in the present paper we limit ourselves to the case summarized in
Table~\ref{tab:1}.

\section{Entanglement entropy}
\label{sect:entanglement_entropy}

In the framework of AdS/CFT correspondence, a holographic method to calculate the entanglement entropy has been proposed in ref.~\cite{Ryu:2006bv}. Following ref.~\cite{Ryu:2006bv}, for a conformal field theory (CFT) which has a dual gravitational configuration living in one higher dimension, the entanglement entropy of the CFT in a subsystem $\mathcal{A}$ with its complement can be obtained by searching the minimal area surface $\gamma_\mathcal{A}$ extended into the bulk with the same boundary $\partial\mathcal{A}$ of $\mathcal{A}$. That is, the entanglement entropy of $\mathcal{A}$ with its complement is given by the ``area law"
\begin{equation}\label{law}
S_\mathcal{A}=\frac{2\pi}{\kappa^2} Area(\gamma_\mathcal{A}).
\end{equation}

\begin{figure}[h]
\centering
\includegraphics[scale=0.6]{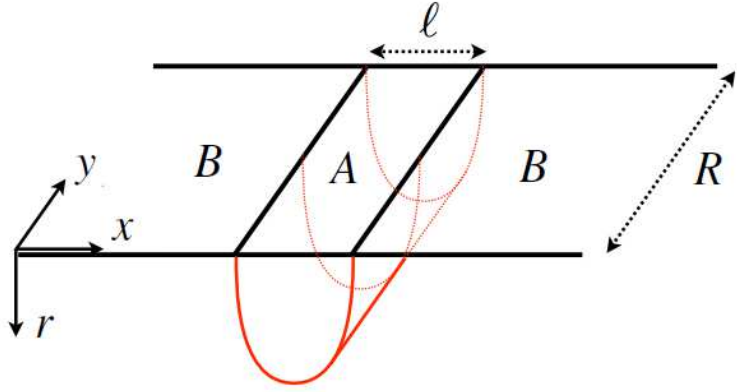}\ \ \ \
\caption{\label{surface}{The minimal surface $\gamma_A$ corresponding to the strip region $\mathcal{A}$ in the boundary. The quantity $\ell$ sets the size of region $\mathcal{A}$.}} 
\end{figure}

In this section we will study the behavior of entanglement entropy in this holographic p-wave superconductor model. We  will consider a belt geometry with a finite width $l$ along the $x$ direction and extends in $y$ direction. The holographic dual surface $\gamma_A$ is defined as a two-dimensional surface
\begin{equation}\label{embed}
t=0,\ \ r=r(x),\ \ -\frac{R}{2}<y<\frac{R}{2}\ (R\rightarrow\infty).
\end{equation}
$R$ is the regularized length in $y$ direction. To avoid the UV divergence, we consider the subsystem $\mathcal{A}$ sits on the slice $r=\frac{1}{\epsilon}$ with $\epsilon\rightarrow 0$ the UV cutoff. More specifically, the holographic surface $\gamma_A$ starts from $x=\frac{\ell}{2}$ at $r=\frac{1}{\epsilon}$, extends into the bulk until it reaches $r=r_*$, then returns back to the AdS boundary $r=\frac{1}{\epsilon}$ at $x=-\frac{\ell}{2}$.  The smooth configuration is shown in figure~\ref{surface}.

Note that there might also exist a piece-wise configuration for which the dual surface goes straight down from the boundary to the horizon, see, e.g., figure 1 in ref.~\cite{Faraggi:2007fu}. However, it has been suggested that generically the smooth one would always dominate at finite temperature involving black hole horizons~\cite{Faraggi:2007fu}. So following the previous studies~\cite{Albash:2012pd,Cai:2012nm,Arias:2012py}, we only consider the case with smooth configuration.

The entanglement entropy of such a subsystem is given by
\begin{equation}\label{entanglement}
S_{\mathcal{A}}=\frac{4\pi L}{\kappa^2}R \int^{\frac{1}{\epsilon}}_{r_*}\frac{r^3}{\sqrt{(r^4-\frac{h(r_*)}{h(r)}r^4_*)f(r)}}dr=\frac{4\pi L}{\kappa^2}R(\frac{1}{\epsilon}+S_E),
\end{equation}
where the UV divergence part $1/\epsilon$ has been separated from the total entropy. Thus $S_E$ is the finite part of physical relevance. The width $l$ of the subsystem $\mathcal{A}$ and $r_*$ are connected by the relation
\begin{equation}\label{width}
\frac{l}{2}=\int_{r_*}^{\frac{1}{\epsilon}}\frac{dx}{dr}dr=\int_{r_*}^{\frac{1}{\epsilon}}\frac{1}{\sqrt{\(\frac{r^6 h^2(r)}{r_{*}^4h(r_*)}-r^2 h(r)\)f(r)}}dr.
\end{equation}
We are here interested in the finite part $S_E$ for the entanglement entropy, which
is of physical relevance. With the background solutions $h(r)$ and
$f(r)$ by solving the equations of motion for the model, we can calculate the entanglement entropy
numerically.

\begin{figure}[h]
\centering
\includegraphics[scale=0.96]{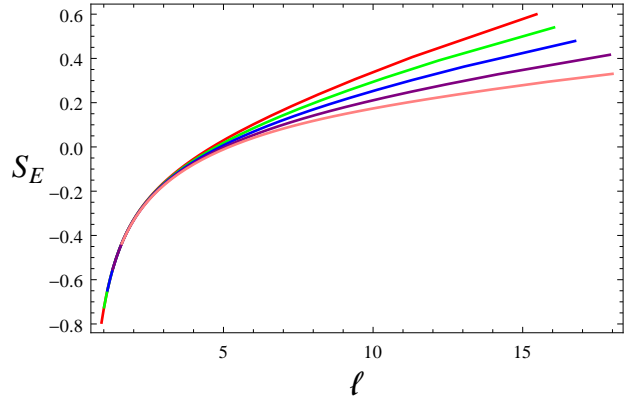}\ \ \ \
 \caption{\label{entanglement_entropy_all} The entanglement entropy as a function of belt width at fixed temperature in the case of $m^2=3/4$ and $q=3/2$. The curves from top to bottom correspond to $T\approx 0.0179,0.0173,0.0165,0.0152,0.0132$, respectively.}
\end{figure}

Our results show that the entanglement entropy $S_E$ with respect to strip width $l$ behaves quite similar for different parameters $m^2$ and $q$. The entanglement entropy versus strip width $l$ is shown in figure~\ref{entanglement_entropy_all} with $m^2=3/4$ and $q=3/2$. In this case, the critical temperature is $T_c \approx 0.0179$~\cite{Cai:2013aca}. Different colorful curves reveal how the entanglement entropy changes with respect to the belt width for different temperatures. Note that in the
present paper we are working in the grand canonical ensemble with fixed chemical potential $\mu=1$. From the top to bottom, the temperature decreases. The curve at the top is for the case with the critical temperature $T_c$, which is identical with the case of the normal phase. From figure~\ref{entanglement_entropy_all}, we can see  that the slope of the curves decreases when the temperature lowers in the superconducting phase. This is expected because the lower the temperature is, the more the degrees of freedom are condensed.

For a fixed temperature, say $T\approx 0.0165$, the blue curve
describes a monotonically increasing behavior of the entanglement
entropy with respect to the belt width. The entanglement entropy is
dominated by the connected surface $\gamma_A$. The wider belt width
corresponds to a larger area of the holographic surface. This
dependence of the entanglement entropy on $l$ is non-trivial. More
explicitly, for large $l$, we see that $S_E$ changes linearly with
$l$. This is because in large $l$ limit, the main contribution of
the integration in \eqref{entanglement} and \eqref{width} to $S_E$
comes from the region near $r=r_*\sim r_h$. Near the horizon, we can
deduce the linear relation $S_E\sim l$. In contrast, the value of
$S_E$ seems to be power-law divergent as $l$ goes to zero.

As shown in Table~\ref{tab:1}, the phase structure is quite
different for the $m^2 \ge m_c^2$ case and $m^2<m_c^2$ case.
In the following we will explore the  behavior of the entanglement entropy
quantitatively for these two cases, respectively.

\subsection{The case with $m^2\geqslant m_c^2$}
In this case we take $m^2=3/4$ as a typical example. For other values of the mass, the behavior of
entanglement entropy is the same qualitatively. Note that in this case, the critical charge $q_c \approx 1.3575$~\cite{Cai:2013aca}.
\begin{figure}[h]
\centering
\includegraphics[scale=0.85]{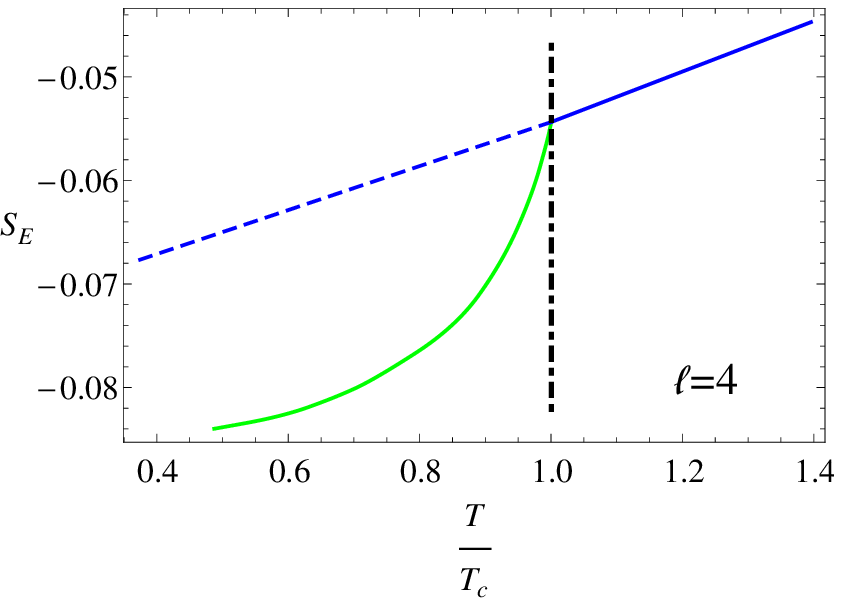}\ \ \ \
\includegraphics[scale=0.85]{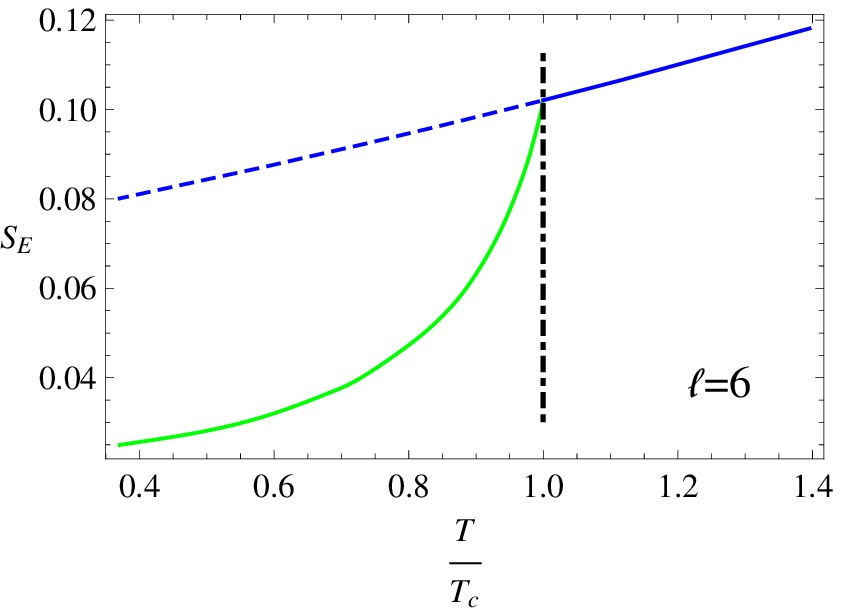}\ \ \ \
 \caption{\label{entropy_1} The entanglement entropy as a function of temperature with fixed belt width in the case $m^2=3/4$ and $q=3/2$ (left plot for $\ell=4$ and right plot for $\ell=6$ ). The blue curves correspond to the entanglement entropy in the normal phase, while the green curves are for the superconducting phase. The physical curve is determined by tracing the lower entropy at a given T. The critical temperature is $T_c\approx 0.00342\mu$.}
\end{figure}

Figure~\ref{entropy_1} shows that the entanglement entropy is continuous at critical temperature $T_c$, but its slope is not. This discontinuity signals a significant reorganization of the degrees of freedom of the system. Since there is a condensate generated at the transition point, it is expected that there is a reduction of degrees of freedom. The behavior of the entanglement entropy versus temperature indicates a typical second order phase transition.

\begin{figure}[h]
\centering
\includegraphics[scale=0.85]{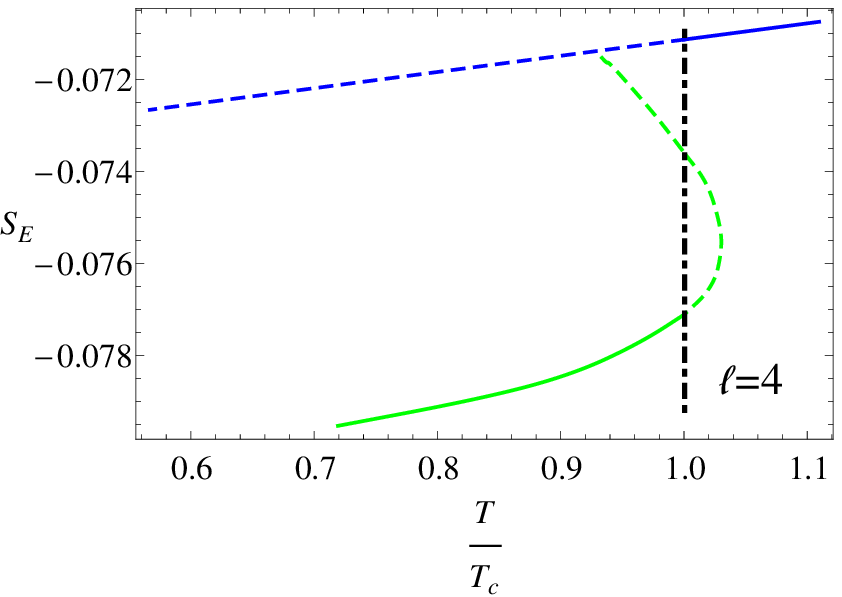}\ \ \ \
\includegraphics[scale=0.85]{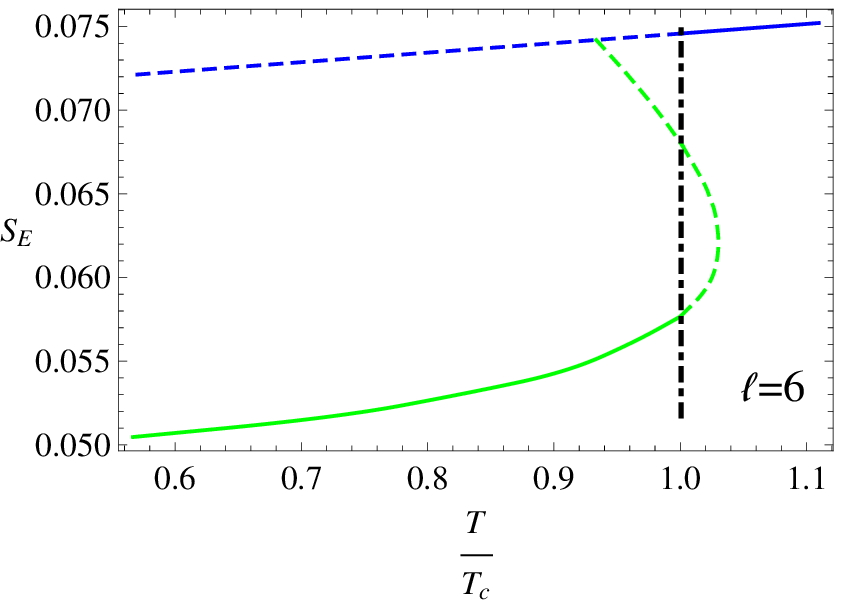}\ \ \ \
 \caption{\label{entropy_11} The entanglement entropy as a function of temperature with fixed belt width in the case $m^2=3/4$ and $q=6/5$ (left plot for $\ell=4$ and right plot for $\ell=6$). The blue curves are from the AdS Reissner-Nordstr\"om solutions, while the green curves are from superconductor solutions. The physical curve is selected by choosing the solid blue curve above the critical temperature $T_c\approx0.00342\mu$, denoted by the vertical dash-dotted line, and the curve which has the lowest entropy below $T_c$.}
\end{figure}

As we decrease $q$ to $q=6/5$ which is less than $q_c\approx1.3575$, the entanglement entropy as a function of temperature is presented in figure~\ref{entropy_11}. In such a case, the entanglement entropy has a jump from the normal phase to the
 superconducting phase at the critical temperature, showing the behavior of a first order phase transition, which is also consistent with the behavior of the thermal entropy. This example shows that entanglement entropy indeed can tell us the appearance of phase transition and its order.

From figure~\ref{entropy_1} and figure~\ref{entropy_11}, we can also see that the behavior of entanglement entropy is the same qualitatively for the belt widths $l=4$ and $l=6$.  The only
difference is that the value of the entanglement entropy for larger
$l$ is larger at the same temperature $T$. This is
consistent with the observation in figure~\ref{entanglement_entropy_all}.

Note that the behavior of the entanglement entropy observed here is qualitatively the same as the one of the thermal entropy considered in  the paper~\cite{Cai:2013aca}. Considering the thermal entropy is equivalent to the black hole entropy in the holographic setup, our results indicate that the black hole entropy could be due to the entanglement entropy~\cite{Jacobson:1994iw,Kabat:1995eq,Solodukhin:2006xv,Emparan:2006ni}.

\subsection{The case with $m^2 < m_c^2$}

\begin{figure}[h]
\centering

\includegraphics[scale=0.85]{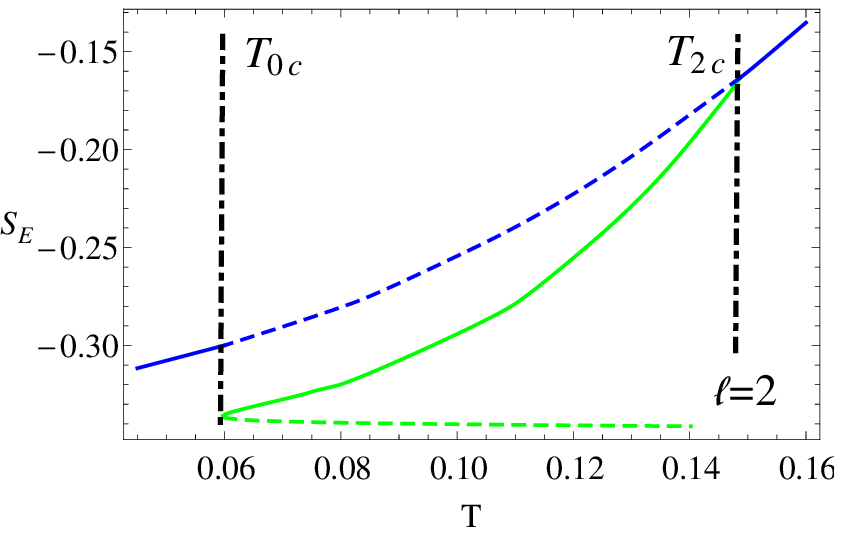}\ \ \ \
\includegraphics[scale=0.85]{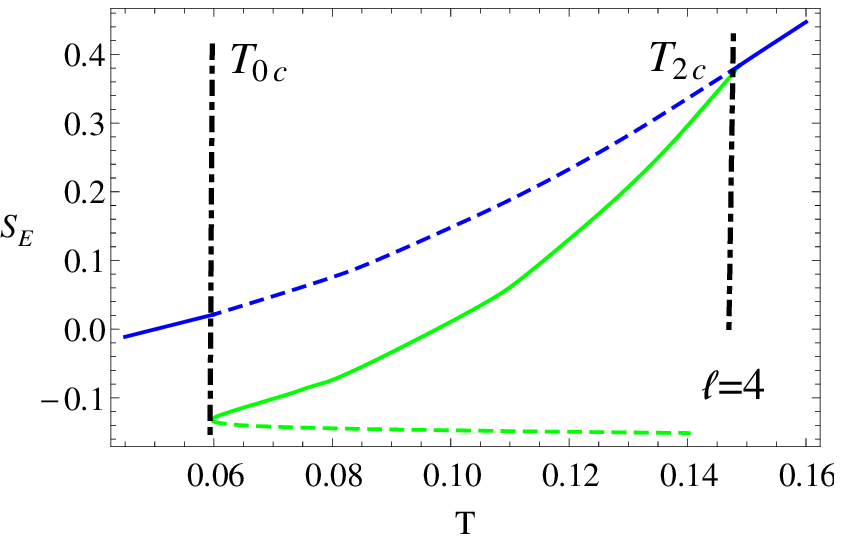}\ \ \ \
 \caption{\label{entropy_3} The entanglement entropy with respect to temperature with fixed belt width in the case $m^2=-3/16$ and $q=2$~(left plot for $\ell=2$ and right plot for $\ell=4$). The blue curves are from the AdS Reissner-Nordstr\"om solutions, while the green curves are from superconducting solutions. The physical branch is selected by choosing the solid curve, which shows a 0th order phase transition at $T_{0c}$ and a 2nd phase transition at $T_{2c}$.}
\end{figure}

\begin{figure}[h]
\centering
\includegraphics[scale=0.85]{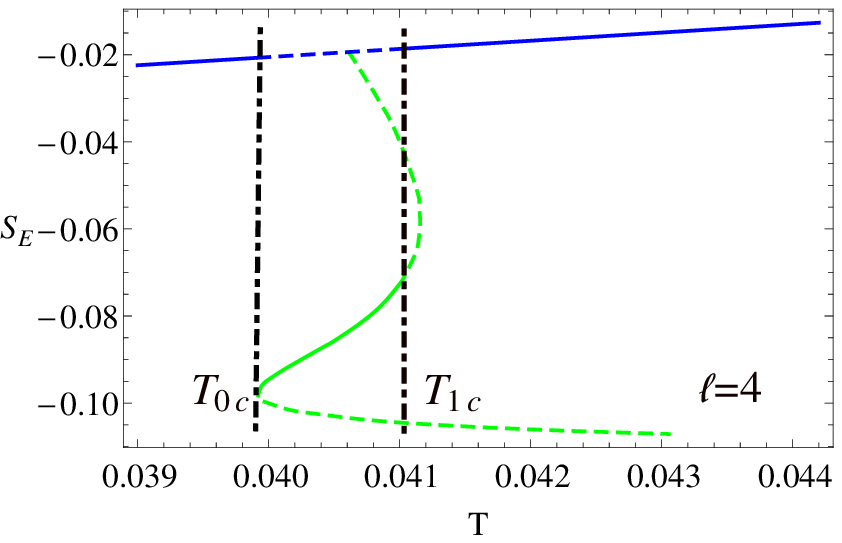}\ \ \ \
\includegraphics[scale=0.85]{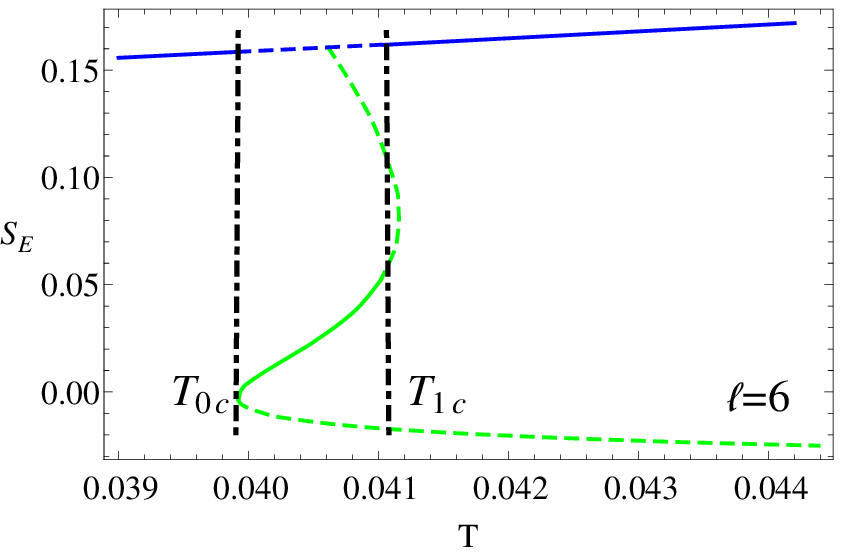}\ \ \ \
 \caption{\label{entropy_4}  The entanglement entropy with respect to temperature with fixed belt width in the case $m^2=-3/16$ and $q=39/40$~(left plot for $\ell=4$ and right plot for $\ell=6$). The blue curves are from the AdS Reissner-Nordstr\"om solutions, while the green curves are from superconductor solutions. The physical branch is selected by choosing the solid curve, which shows a 0th order phase transition at $T_{0c}$ and 1st order phase transition at $T_{1c}$. }
\end{figure}

\begin{figure}[h]
\centering
\includegraphics[scale=0.85]{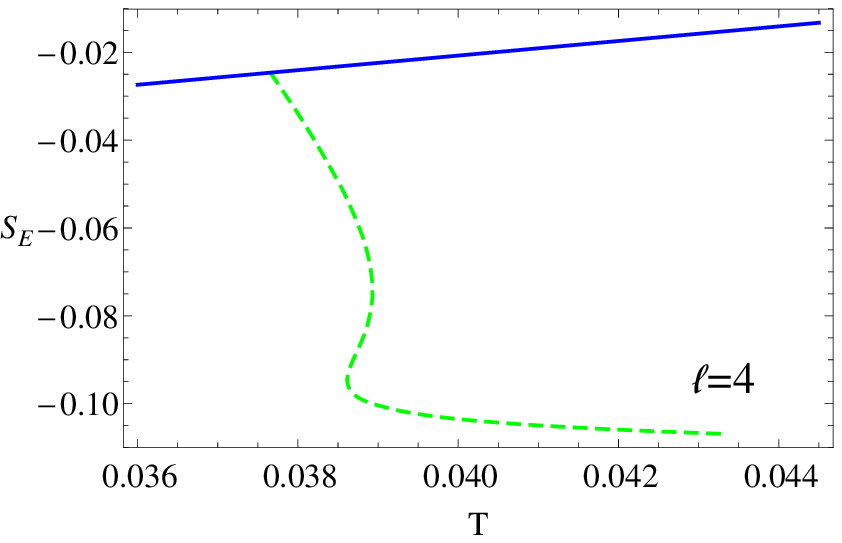}\ \ \ \
\includegraphics[scale=0.85]{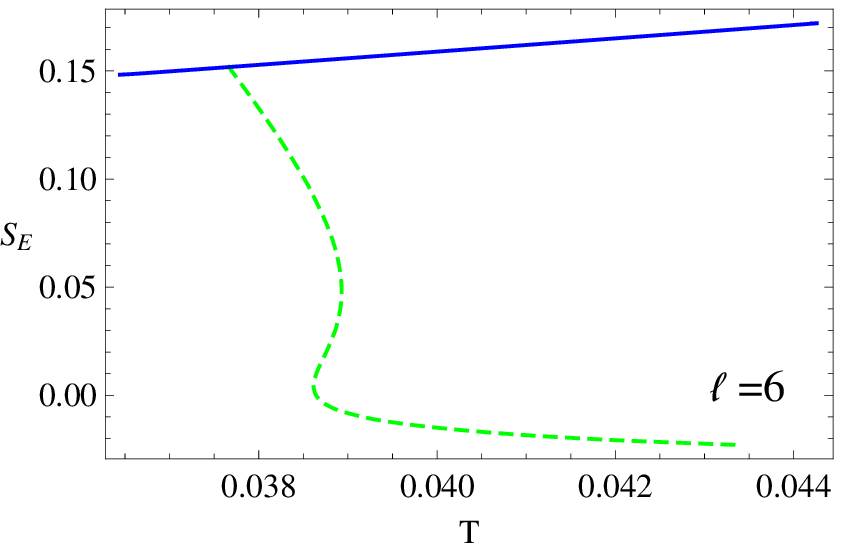}
 \caption{\label{entropy_5} The entanglement entropy as a function of temperature at fixed belt width in $m^2=-3/16, q=19/20$ case (left plot for $\ell=4$ and right plot for $\ell=6$). The solid curves are from the AdS Reissner-Nordstr\"om solutions, while the dashed curves are the ``retrograde condensation" phases. There is no phase transition at all in such case.}
\end{figure}

\begin{figure}[h]
\centering
\includegraphics[scale=0.85]{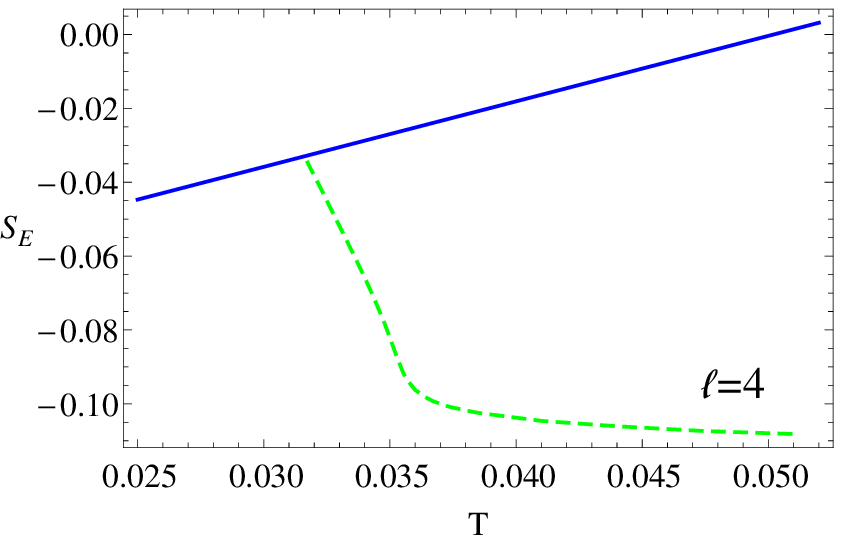}\ \ \ \
\includegraphics[scale=0.85]{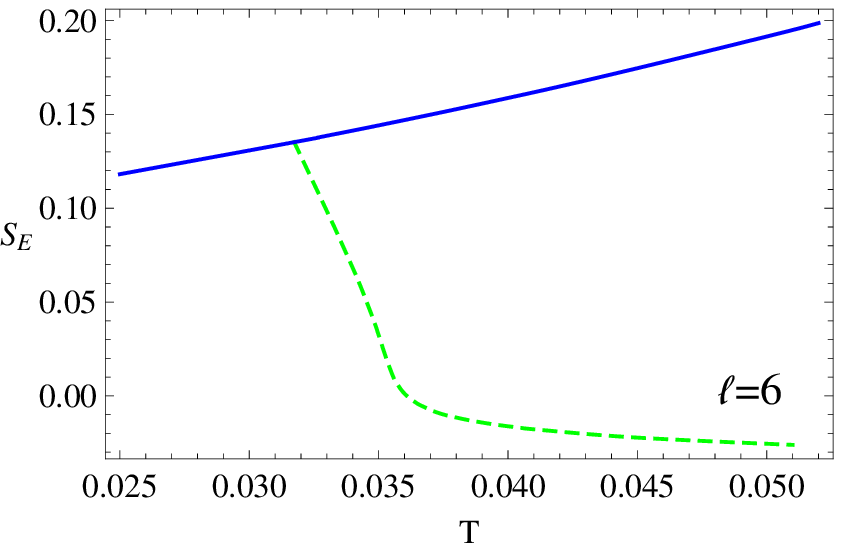}\ \ \ \
 \caption{\label{entropy_6} The entanglement entropy as a function of temperature at fixed belt width in $m^2=-3/16, q=9/10$ case (left plot for $\ell=4$ and right plot for $\ell=6$). The solid curves are from the AdS Reissner-Nordstr\"om solutions, while the dashed curves are from the ``retrograde condensation"  phases. There is no phase transition at all in such case.}
\end{figure}

In this case we take $m^2=-3/16$ as a typical example (The BF bound is still satisfied in this case). Note that in this case the critical charges are $q_{\alpha} \approx 1.0175$ and $q_{\beta} \approx 0.9537$~\cite{Cai:2013aca}, respectively. We find different behaviors of entanglement entropy for the cases $q \ge q_{\alpha}$,
$ q_{\alpha} >q> q_{\beta}$, and $ q \le q_{\beta}$, respectively.
In figure~\ref{entropy_3}-\ref{entropy_6} we plot the entanglement entropy versus temperature for the cases of $q=2$, $q=39/40$, $q=19/20$ and $q=9/10$, respectively.

In figure~\ref{entropy_3} we plot the entanglement entropy versus temperature for $q=2$ as an example
in the case of $q>q_{\alpha}$.  We can see that at the second order transition point $T=T_{2c}$, the entanglement entropy
is continuous, but its slope is not, while at the zeroth order transition point $T=T_{0c}$, the entanglement entropy has a
jump. In addition, in the superconducting phase, the entanglement entropy has two branches: the solid green curve is physically
favored, while the dashed one is not physically favored since for the latter, the solution has a higher free energy than the
hairy black hole solution for the former.

For the case $q_{\beta}<q<q_{\alpha}$, we consider the example with $q=39/40$. The
behavior of the entanglement entropy as a function of temperature is
shown in figure~\ref{entropy_4}. We can see from the figure that
the entanglement entropy has a jump both at the first order transition point at $T=T_{1c}$
and at the zeroth order transition point at $T=T_{0c}$. But the two situations are different. For the
former, both the entanglement entropies in the normal phase and superconducting phase can be connected by a
metastable superconducting phase (dashed curve), while for the latter, there does not exist such a metastable phase.

For the case $q \le q_{\beta}$, we consider two examples of $q=19/20$ and $q=9/10$. Their behaviors of entanglement entropy are shown in
figure~\ref{entropy_5} and figure~\ref{entropy_6}, respectively. In this case, the condensation appears in the high temperature regime, and hairy black hole solutions have higher free energy than the RN-AdS black hole at the same temperature, therefore the hairy black hole solutions are not physically favored (bottom left and bottom right plots in figure~\ref{freen}). We see from the figure that the entanglement entropy $S_E$ is
multi-valued for $q=19/20$, while it monotonically decreases when the temperature increases for $q=9/10$, for the hairy black hole solutions.

Once again, we find that the behviors of the entanglement entropy are the same qualitatively for different belt
widths. Comparing the behaviors of the entanglement entropy with corresponding ones of thermal entropy studied
in~\cite{Cai:2013aca}, we find that both the behaviors are the same qualitatively. This further supports that
entanglement entropy is a good measure of degrees of freedom for quantum field theory, and also it tells that entanglement
 entropy is a good probe to phase transition and various phases in holographic superconductor models.

\section{Conclusions and discussions}
\label{sect:conclusion}

In  recent works~\cite{Cai:2013pda,Cai:2013aca,Cai:2014ija} we have constructed a
holographic p-wave superconductor model in a four dimensional
Einstein-Maxwell-complex vector field theory with a negative
cosmological constant. Depending on the mass square $m^2$ and the charge $q$ of the
complex vector field $\rho_{\mu}$, the model exhibits a rich phase
structure. The second order, first order and even zeroth order phase transitions would
appear in this model and  the so-called ``retrograde condensation" also could happen in some
parameter space.  In this paper we have continued to study this model by investigating the behavior
of holographic entanglement entropy for a belt geometry at the AdS boundary.

The main results of this paper can be found from figures~\ref{entanglement_entropy_all} to \ref{entropy_6}.
In figure~\ref{entanglement_entropy_all} we show the behavior of the entanglement entropy with respect to the belt width for
different temperatures in the superconducting phase. For the case with a fixed temperature, the entanglement entropy increases
with the belt width, while in the case for a fixed width, it decreases when the temperature is lowered. This is expected because
entanglement entropy is a measure of degrees of freedom for the dual field theory and when the temperature lowers, more degrees of
freedom will be condensed.

Figures~\ref{entropy_1}-\ref{entropy_11} and \ref{entropy_3}-\ref{entropy_6} show the behavior of the entanglement entropy with respect to temperature in various
phases, which depend on the model parameters $m^2$ and $q$. We have observed that at the second order phase transition point, the
entanglement entropy is continuous, but its slop is discontinuous, while at the first order and the zeroth order transition points,
the entanglement entropy has a jump. But for the latter two cases, the situations are different. For the first order phase transition,
the two entanglement entropies for the normal phase and the superconducting phase can be connected by a metastable condensed phase, while
in the case of the zeroth order phase transition, there does not exist the metastable condensed phase. Here we remind again
that the zeroth order phase transition discussed in this paper might be not physical, as we mentioned in the above, because there might
exist more stable non-trivial black hole solutions in this model.

In this present study, we have seen that the behavior of entanglement entropy always shares some similarity with thermal entropy~\cite{Cai:2013aca}. The similarity can be understood as the fact that the entanglement entropy is contaminated by the thermal fluctuation at finite temperature, and the entanglement entropy approaches to the thermal entropy at high temperatures. In this sense, the entanglement entropy is a unique order parameter one can use even at zero temperature and it deserves to further investigate the behavior in the holographic superconductor/insulator transition in the Einstein-Maxwell-complex vector model. Indeed, the entanglement entropy in some superconductor/insulator models at zero temperature were studied and some interesting behaviors were uncovered~\cite{Cai:2012sk,Cai:2012es,Yao:2014fua}.

Note that in this paper we  have only studied the entanglement entropy with
finite width along the $x$ direction. In such a model, the spatial
rotational symmetry is broken in the superconducting phase. To
capture the anisotropy in $x$ direction and $y$ direction, ones
should also study the entanglement entropy along $y$ direction. We
do not calculate it here because the behavior of the entanglement
entropy along $y$ direction is expected to be qualitatively similar
to the case in $x$ direction. As a non-local quantity, the
entanglement entropy describes the new degrees of freedom emerging
in the superconducting phase. Since the degree of anisotropy of the
superconducting phase is characterized by the value of condensate,
therefore, it is not expected to have qualitative difference for entanglement entropy
between along $x$ direction and $y$ direction~\cite{Cai:2013oma}.

Furthermore, apart from the entanglement entropy, there is another nonlocal
quantity, Wilson loop, which can describe some aspects  of
gauge field theory. The study of such a quantity can also shed some lights
into the holographic superconducting phase transition. We
expect to report further progress on these issues in
future.

\section*{Acknowledgements}

This work was supported in part by the National Natural Science Foundation of China (No.11035008, No.11375247, No.11205226 and No.41231066), and in part by the Ministry of Science and Technology of China under Grant No.2010CB833004. L.F.L and C.S would like to appreciate the National Basic Research Program of China (973 Program) Grant No.2011CB811404 and the Specialized Research Fund for State Key Laboratories.

\end{document}